\documentclass[letterpaper]{jpconf}
\usepackage{graphicx}
\begin{document}

\newcommand{\diffunit}{$\mathrm{GeV\;cm^{-2}\;s^{-1}\;sr^{-1}}$}
\newcommand{\pointunit}{$\mathrm{TeV\;cm^{-2}\;s^{-1}}$}
\newcommand{\dNdE}{E^{2}_{\nu} \times dN_{\nu}/dE_{\nu}}
\newcommand{\Nch}{$N_{\mathrm{ch}}\;$}
\newcommand{\ea}{{\it et al} }
\newcommand{\ic}{IceCube}
\newcommand{\esqdnde}{$\mathrm{E^{2}_{\nu} \times dN_{\nu}/dE_{\nu}}$}
\newcommand{\puneicrc}{2005 Proc. 29th Int. Cosmic Ray Conf., Pune}
\newcommand{\ar}{Ahrens J {\it et al} }
\newcommand{\am}{Ackermann M {\it et al} }
\newcommand{\ab}{Achterberg A {\it et al} }

\title{Neutrino astronomy with IceCube and AMANDA}

\author{Gary C. Hill, for the IceCube collaboration \cite{icecube}}

\address{Department of Physics, University of Wisconsin, Madison}

\ead{ghill@icecube.wisc.edu}


\begin{abstract}
Since the early 1990s, the South Pole has been the site of the construction of the 
world's first under-ice Cherenkov neutrino
 telescopes - AMANDA and IceCube. The AMANDA detector
was completed in 2000, and its successor
 IceCube, a kilometre scale neutrino detector, began construction in 2005. 
 Completion of IceCube is scheduled for
2011. This paper will give an overview of the history, construction, latest 
physics results and potential
of these detectors. 
\end{abstract}

\section{The appeal of neutrino astronomy}
The road to a kilometre scale neutrino detector, pioneered by the DUMAND 
collaboration, has seen the operation of the first generation experiments,
AMANDA and Lake Baikal, as well as initial construction and planning for IceCube,
ANTARES, NESTOR, NEMO and KM3NET. 
The discovery  of neutrinos with these detectors will  hopefully  extend and complement the
knowledge of the universe to date gained through cosmic ray and gamma ray observations.
While the nature and location of the cosmic ray sources are unknown, there are many confirmed
sources of TeV gamma-rays.
If one of these turned out to also be a neutrino source, then
a hadronic accelerator central engine might be driving cosmic ray, gamma and neutrino
production \cite{PR}.

A neutrino detector like IceCube or AMANDA uses an array of photomultipliers to
record Cherenkov light from through-going muons, or from point-like shower (``cascade'')
events. Muons result from charged current interactions of neutrinos in the detector 
volume, or in the surrounding ice and rock. Cascade events result from charged and
neutral current interactions of all neutrino flavours. 

The backgrounds to a search for a flux of high-energy extra-terrestrial neutrinos
at the earth are atmospheric muons and neutrinos from the interaction of cosmic rays in the 
earth's atmosphere. 
 The atmospheric muons are eliminated by looking for events moving
upward through the detector -- only neutrinos can penetrate the earth. A small fraction of
the large downgoing muon  flux  will be falsely reconstructed in the upward direction.
These are removed by  tight requirements on the fitted track. 
  After atmospheric muons are eliminated, there is a flux of atmospheric
neutrinos
 seen in a detector. This can be used as a calibration test beam to check
the understanding of the detector, or be used to look for new neutrino physics. 
 A search for point sources of neutrinos is made by looking for an excess of events from
a direction in the sky. Electromagnetic observations by other 
 detectors may provide information to reduce the
time over which such a search is made - for instance in a search for neutrinos
correlated with a gamma-ray burst. 
 One can also look for a diffuse excess of neutrinos from the sum of all sources in the 
universe. Since the extra-terrestrial flux predictions tend to go as $dN/dE \sim E^{-2}$,
 one 
looks for higher energy events in the detector to separate them from the more
steep atmospheric neutrino spectrum ($dN/dE \sim E^{-3.7}$). 
\section{Physics results from AMANDA}
The first detection of muon Cherenkov radiation
 in polar ice was made in Greenland in 1990 \cite{muons91}, 
using three photomultipliers deployed to a depth of about 200 metres. 
Following this success, similar tests were made at the South Pole over
the next years, with the AMANDA-A detector deployed in 1993-94 \cite{ice1}. 
Construction of the presently operating 
AMANDA detector took place from 1995 to 2000, over which
time 677 optical modules were deployed over 19 strings, to 
depths ranging from 1500 to 2000 metres. The properties of the polar ice, critical
for understanding of the detector, have been measured using light sources
in the array \cite{icepaper}. Although most of AMANDA used analogue signal technology,
 digital technology, eventually chosen for IceCube, was
tested on one string \cite{string18}.

\subsection{Atmospheric neutrinos}
While three neutrino candidates were observed with the first four strings of AMANDA \cite{AMA-B4},
the first compelling evidence of high-energy atmospheric neutrinos came from the 
10 string 1997 data set, where 16 upgoing events were left after data reduction \cite{firstatmos}. Dramatic
improvements in the analysis techniques \cite{recoAMANDA} increased this number to about 300 \cite{atmosnu1997,atmos-nature}.
Over the entire life of AMANDA-II, many thousands of atmospheric neutrinos
have now been observed \cite{ackermann-ps,ps2000-04}. These are the highest energy 
neutrinos ever observed. 
 The observed rate is consistent with the uncertainties in
 theoretical
predictions \cite{bartol2004,honda2004}. A regularised unfolding technique has been used to make a 
best-fit to the originating energy spectrum; again consistency with expectation is 
seen \cite{Munich}. The agreement of the atmospheric neutrino measurements with
expectations shows that the detector is working as expected. 

\subsection{Point sources}
 Several searches for northern hemisphere
point sources of neutrinos have been conducted with the 
AMANDA detector, for the  1997 \cite{ps-1997}, 2000 \cite{ps-2000}
 and 2000-02 \cite{ps-2000-02} data sets. The most recent search used data from 2000-04, corresponding,
after correction for down-time of the detector, to 1001 days of live time \cite{ackermann-ps,ps2000-04}.
 The
final event set consists of 4282 upward moving events, believed to be
atmospheric neutrinos. 
Several  search methods were used  to look for point sources in the northern sky. 
For each, the expected background for any source is found from off-source data
from the same declination band. The expected sensitivity is found from 
simulations of neutrino interactions, muon propagation, and 
 the full detector response to the Cherenkov light emitted. 
Full-sky searches (looking for a hot spot anywhere in the sky), specific 
source searches, and stacking searches were conducted. 
 The full-sky and specific source searches were optimised in an unbiased fashion
  to produce
the best limit setting potential \cite{mrp}. 
The 90\% confidence level 
sensitivity of the 
full-sky search to an $E^{-2}$ flux (assumed to have a $\nu_{\mu}:\nu_{\tau}$ ratio of 1:1), relatively constant with declination, is about
\esqdnde $<  10^{-10}$ \pointunit. The numbers of observed 
events across the sky were consistent with the background expectations, leading to the
same result for  the average all-sky experimental
limit.  The highest significance seen was 3.7$\sigma$ and, via scrambled
random sky maps, the probability of seeing something this significant
or higher was found to be 69\%. 
Searches for 32 specific candidate sources, and searches made where the events from
 objects belonging to
 common classes were summed, were made. 
Limits were placed on the neutrino fluxes from the objects \cite{Gross-stacking,ps2000-04}. 
 For a source above the horizon, SGR 1806-20, a search for muons from
both neutrinos and gamma-rays was made. With no  significant signal seen,
 limits were placed on the gamma and neutrino fluxes from the source \cite{sgr}.
   While not truly a point source, the galactic plane was searched for an excess of 
neutrinos from cosmic ray interactions with the dust, using similar methods as employed in
 the 
point source searches.
No excess
of events was seen and limits on models were set \cite{Kelley}.

\subsection{GRBs}
Gamma-ray bursts are some of the most energetic phenomena in the universe, with
emission timescales
as short as  seconds. During the life of AMANDA, satellites such as the CGRO, with the
BATSE detector, and the IPN satellites, including HETE and Swift, have recorded
gamma emissions from many GRBs. 
Waxman and Bahcall theorised that GRBs may be the source of the highest energy
cosmic rays \cite{WB97}. In this ``fireball'' model, neutrinos would also be produced. 
The AMANDA data has been searched for neutrinos in spatial and temporal 
coincidence with about 400 GRBs \cite{Kuehn}. The addition of a time cut on the search 
greatly reduces the expected background  to of order one event over the sum of
all GRBs searched.  No event has been observed in coincidence with
a GRB, consistent with  this small total expected background.  Limits on 
the fluxes from all bursts, classes of bursts,  and individual bursts, have
been placed. The limits from all bursts are within a factor 4 of the Waxman-Bahcall prediction.
 In another analysis, the observations from each individual
 burst are 
interpreted in light of all information known about that burst from other 
wavelengths, via an individually calculated neutrino flux. An
analysis of this type has been 
performed for GRB030329 \cite{Stamatikos}.  The study of further GRBs is in progress. 
 Searches for cascade like events from GRBs  have been made \cite{Hughey}.
 All-time and rolling time
window searches have been performed and limits placed on models of neutrino production.

\subsection{WIMPs}
The mystery of the dark matter, responsible for some 23\% of the 
energy density of the universe, is a target of the search for WIMPs
 (Weakly Interacting Massive Particles) with AMANDA.
A likely dark matter candidate is the neutralino - the lightest 
supersymmetric particle in in most supersymmetric extensions of the 
standard model. 
 After some time, these
would become gravitationally trapped in the centre of the earth and
sun, where they could pair-wise annihilate via several paths to produce
neutrinos. Thus, AMANDA searches for excesses of 
neutrinos from the centre of the earth (1997-99 data \cite{WIMPS97earth,WIMPs97-99earth}),
 and from the sun (2001 data \cite{WIMPs2001sun}).
To date, neither the earth nor sun has been revealed as an annihilation
site for neutralinos, and these non-observations  place bounds on various
parameters in the supersymmetric extensions of the standard model.
 Once
all current data is analysed, these bounds  will be competitive and 
complementary with those from direct 
 detection experiments like CDMS.

\subsection{Diffuse searches}
\label{diffuse}
To search for a diffuse flux of neutrinos from the sum of sources in the 
universe, one must look for neutrinos in excess of the expectation for
  atmospheric neutrinos. The extra-terrestrial flux is expected to  have
a harder spectrum ($\sim E^{-2}$) than the atmospheric neutrinos ($\sim E^{-3.7}$), 
so searches are designed where event energies are estimated.
 Three types of diffuse search are conducted with AMANDA, one sensitive to muon-neutrinos,
and the other two sensitive to all flavours. The muon search seeks to isolate muon
tracks and use event observables related to the energy.
  One style of all-flavour search focuses on cascade-like events - and is thus
sensitive to neutral and charged current interactions of all flavours.
   Cascades from charged current interactions come from electron and tau neutrinos, and 
from some muon-neutrinos where most of the energy goes into the cascade, leaving only a 
short track from a low energy muon. These searches are mostly sensitive to cascade events
 contained in the detector volume.
  The second type of all-flavour search looks for large cascade and muon events from
extremely high energy neutrino interactions, including events where the cascade or
muon is well outside the volume of the detector. Due to attenuation of neutrinos in the
earth,  these searches are most sensitive to
horizontal events, with the main background being energetic cosmic ray muon bundles.
  
  Unlike a point source search, a diffuse search strictly has no 
 ``off-source'' region where
data can be used to estimate the background. Thus the analysis relies on 
theoretical predictions of the atmospheric neutrino fluxes for background
estimations. In practice, the observed lower energy events are  used to 
 place some   constraint on the atmospheric models before they are used to estimate
the high energy background. As for other analyses, downgoing muons are used as a 
calibration beam to check that the  detector would be sensitive to the types of
high-energy events expected from extra-terrestrial neutrinos.

 Table \ref{diffusetable} summarises the results of the different searches for
a diffuse flux of neutrinos with the AMANDA data sets, taken from 1997 to 2003. 
In the results reported here, the all-flavour analyses assume a 1:1:1 electron,
muon and tau flavour mixture at the earth, due to maximal neutrino 
oscillations. These limits can be converted (and compared) to muon limits by 
dividing by three.
 
  Two \emph{all-flavour cascade} searches have been performed,
 on the 1997 \cite{1997cascade} and
2000 \cite{2000cascade} data sets.
 The limit for the 2000 data  improved by an order of magnitude
over that for 1997. In a similar energy range ($20-5\times10^{4}$ TeV),
 the Baikal collaboration has
recently analysed  1038 days (1998-2003) of 
data from  the NT-200 experiment, leading to a slightly better limit of
 \esqdnde $= 8.1\times10^{-7}$ \diffunit \cite{Baikal98-03}.

 At higher energies, these data sets have been analysed with the  \emph{all-flavour UHE} 
method \cite{1997UHE,Gerhardt-ICRC,Gerhardt-SUSY}.
 Although the sensitivity of the 2000 search ( \esqdnde $= 3.7\times10^{-7}$ \diffunit) was
improved over 1997,  the experimentally obtained limit for 2000 turned out to be the same as that
for 1997, due to the observation of
 a non-significant excess of events.
These limits are the best of any detector at energies up to $\sim 1$PeV.

\begin{figure}
\centering
\includegraphics*[width=0.85\textwidth]{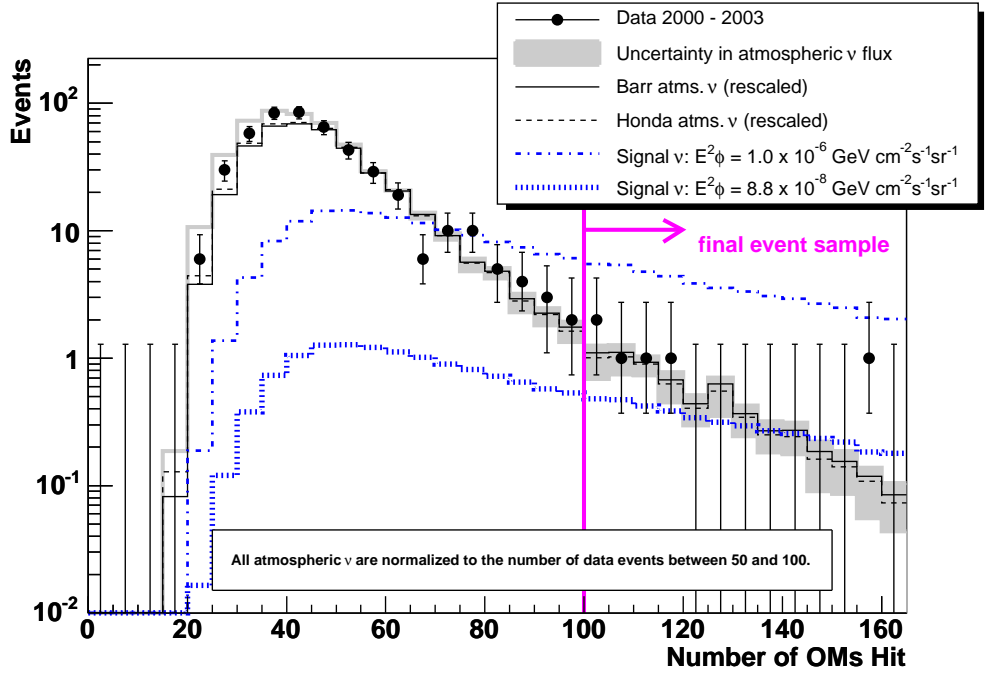}
\caption{\label{nch_sigrescaled} \Nch, the number of OMs triggered, for the AMANDA 2000-03 muon track
diffuse analysis. The data 
is compared to atmospheric neutrino expectations \cite{bartol2004,honda2004}. The signal prediction for
an $E^{-2}$ flux is rescaled to reflect the event limit derived
from the  background and events above \Nch$=100$.}
 \end{figure}

Searches for a diffuse flux, using reconstructed contained muon events, have been
made on the 1997 \cite{97diffuse}, 2000 and 2000-03 data sets. For the year
 2000 data set, a regularised
unfolding of the energy spectrum was conducted. This spectrum was statistically compared
with the atmospheric neutrino expectation and a limit on a diffuse $E^{-2}$ flux
derived \cite{Munich}. For the 2000-03 data, the muon analysis used the
number of optical module channels per event that reported at least one Cherenkov
photon (\Nch$\!$) as an energy estimator. The harder expected extra-terrestrial flux
would produce a flatter \Nch distribution than that for atmospheric neutrinos (see figure
\ref{nch_sigrescaled}).
 Before looking at the data, an optimal cut of \Nch was found in order to produce
the best limit setting sensitivity of the search \cite{mrp,hodgesICRC}.
 The data above this cut (\Nch $> 100$)
were kept blind while the lower \Nch events were compared to atmospheric neutrino 
expectations. The Bartol \cite{bartol2004} and Honda \cite{honda2004}
 atmospheric neutrino fluxes
were varied to account for systematic uncertainties, then constrained by normalisation with the
low \Nch data. The remaining spread in the high \Nch region was used to calculate
an error on the expected number of events above the \Nch $> 100$ cut. The results are
shown in figure \ref{nch_sigrescaled} and compared to the data. Above the cut, 6 events
were seen, where 6.1 were expected. Using the range of atmospheric uncertainty (shaded
band in figure \ref{nch_sigrescaled}) in the limit calculation \cite{ch} 
leads to a limit on an $E^{-2}$ flux of
muon-neutrinos, at the earth, of
 \esqdnde $= 8.8\times10^{-8}$ \diffunit. This limit is valid in the energy range 16-2500 TeV
 and 
is the best limit
of any neutrino detector to date. Limits were also placed on specific extra-terrestrial
models and on the flux of prompt, charm-meson neutrinos from the earth's atmosphere \cite{hodges-tev}.

\begin{table}
\caption{{\bf Summary of AMANDA diffuse neutrino flux results, 1997-2003}.
The results labelled ``muon'' are for analyses sensitive to neutrino-induced
muon tracks in the detector, and give limits on the muon-neutrino flux at
earth. The ``all-flavour'' analyses are sensitive to
events from muon, electron and tau neutrinos, and place limits on the
total neutrino flux at the earth, assuming a 1:1:1 flavour ratio  due
to maximal mixing neutrino oscillations during propagation to the earth.
Assuming this 1:1:1 flavour ratio,
the muon-neutrino limits may be
 converted to all-flavour limits
 by multiplying by three.}

\label{diffusetable}
  \begin{center}
   ~\\*[0.2cm]
    \begin{tabular}{lcccc}
      \hline
      {\bf Data set} & {\bf Detection channel} & {\bf Neutrino energy} & {Limit \esqdnde ($\mathrm{90\%c.l.}$)}   \\
       & &{\bf range TeV} & {\bf \diffunit} & \\
      \hline \hline
 1997 & muon \cite{97diffuse} & $6-10^3$  & $8.4 \times 10^{-7}$  \\
 1997 & all flavour, UHE \cite{1997UHE} & $10^3-3\times10^6$ & $9.9 \times 10^{-7}$  \\
 1997 & all flavour, cascade \cite{1997cascade} & $5-300$ & $98.0 \times 10^{-7}$  \\
 2000 & all flavour, cascade \cite{2000cascade}  & $50-5\times10^3$   & $8.6 \times 10^{-7}$  \\
 2000 & all flavour, UHE \cite{Gerhardt-ICRC,Gerhardt-SUSY}   &$1.8\times10^2-1.8\times10^6$& $9.9\times10^{-7}$  \\
 2000 & muon, unfolding \cite{Munich} & $100-300$ & $2.6 \times 10^{-7}$  \\
2000-03 & muon & $16-2.5\times10^3$ & $0.88 \times 10^{-7}$  \\
\hline \hline
    & & &  \\
                                                                                                                  
       \hline
\end{tabular}
\\*[1.cm]
\end{center}
\end{table}

\subsection{Supernovae, cosmic ray composition, monopoles and new physics}
AMANDA is a supernova detector, with sensitive coverage of our galaxy \cite{sn-97-98}.
 A burst of low energy electron-neutrinos
from a supernova would produce an increase in the rates of all optical 
modules over a short time ($\sim 10\; {\rm seconds}$).
 The AMANDA supernova system is part of
the Supernova Early Warning System (SNEWS). 
AMANDA, in conjunction with the SPASE surface air shower detector, has been 
used to study the composition of cosmic rays near the knee \cite{composition}. 
Searches for magnetic monopoles have been made, and
Lorentz invariance and 
 decoherence are
 two of the ``new physics'' tests being 
conducted with atmospheric neutrino data from AMANDA.

\section{IceCube: The future is now}
\subsection{Construction and Performance}
The first of the next generation kilometre scale neutrino
telescopes, IceCube, will consist of an in-ice cubic kilometre
neutrino detector, and a kilometre square surface cosmic
ray air shower detector (IceTop).
  Construction began
 at the South Pole 
during the austral summer 2004-05, with 1 in-ice string, and
4 IceTop stations deployed \cite{icperf}.
  During the second summer season,
8 more strings and 12 IceTop stations were installed.
The goal is to complete construction in early 2011, with
80 strings (4800 modules) and stations (320 modules)
 completed. The in-ice strings will
instrument a kilometre volume between 1500 and 2500 metres
depth, and the IceTop array will cover a square kilometre
at the surface. The same design of DOM (Digital Optical 
Module) is used throughout the detector. These consist of
pressure spheres containing 10 inch photomultiplier tubes,
the signals of which are digitised inside the module and 
then sent to the surface data acquisition system. The DOMs
differ from the AMANDA modules in that the full time series
of photons (the ``waveform'') is captured. 

The holes are drilled with a hot water system, 
taking about 30 hours to drill to the final depth, then
10 hours to
 ream back up, depositing more energy to leave a hole at
the correct size during the string deployment.
Deployment of a string takes about 12 hours - 8 hours for
module attachment, 4 hours to lower to the final depth. 
IceTop tanks are installed in shallow trenches dug near each
string location, and are filled with water, which is 
allowed to slowly freeze back about the modules, to prevent 
formation of bubbles. 

The deployed hardware has performed up to expectations
to date. Detailed studies of the first string and IceTop
tank behaviour has been published \cite{icperf}. Two upward moving
events were detected with the single string, consistent
with an atmospheric neutrino origin. The presently operating
9 string and 16 station detector is performing well. Upward
moving neutrino events have been seen. Atmospheric muons have
been tracked in the in-ice array. Air showers have been reconstructed
with IceTop, and coincident events, where IceTop sees an air shower and
the in-ice array sees the penetrating muons, have been studied. 
First physics analyses are well underway. 

\subsection{Physics potential}
An initial potential performance study for the 
in-ice array of IceCube was completed before
construction began \cite{icsens}. The simulation and reconstruction 
programs were those used in AMANDA,
adapted to  the larger IceCube detector. 
As such, no usage of the DOM waveform information was made  in
the reconstruction. The assumed flux of charm atmospheric
 neutrinos \cite{rqpm}
was chosen conservatively; if in reality this background
 turns out
smaller, then the predicted sensitivities will be better than
 those quoted. 
 A median angular resolution of better than 1$^\circ$
is seen for muon energies greater than 1 TeV. The effective area for
muon detection exceeds the geometric kilometre area at 10 TeV, rising
to 1.4 square kilometres for events in the 1 to 100 PeV energy range.
The sensitivity to diffuse and point sources of neutrinos has been
estimated. For three to five years of observation, the limit on an $E^{-2}$
flux of diffuse neutrinos would be 
about thirty times smaller
than the AMANDA-II four-year muon limit (section \ref{diffuse}),
 and a flux one-tenth of the AMANDA-II
limit would be detectable at $5\sigma$ significance in that time. 
For point sources, similar results are obtained. For GRBs, the
Waxman-Bahcall flux would be constrained after the observation of
about 100 GRBs, and 500 GRBs would be needed to observe that flux at
a $5\sigma$ significance. 

\section{Conclusions}
  The long-held dream of a large volume, high energy neutrino detector is finally
 a reality at the South Pole. The last decade has been one of 
technology, deployment, and analysis development with the AMANDA detector,
leading to the design and construction of IceCube. IceCube, 
slated for completion in 2011, is already producing physics data, and once
completed, will have unprecedented sensitivity to sources
of extra-terrestrial neutrinos, hopefully leading to new discoveries about the 
nature of the cosmos.

\end{document}